\documentclass[conference]{IEEEtran}
\usepackage{defpaper}
\usepackage{epsfig}
\usepackage{latexsym}

\newcommand{\filefig}[4]{
  \begin{figure}[htb]
  \vskip 1pt
    \begin{center}
      \setlength{\epsfxsize}{#4}
      \leavevmode
      \epsfbox {#2}
      \caption{\protect {#1}}
      \label {#3}
   \end{center}
\end{figure}}

      {
      \begin{minipage}
      {0.462\textwidth}\medskip\noindent
       \scriptsize\raggedright\sf\noindent
       {\bf Input:} #2\\
       {\bf Output:} #3\\
       \rule[1pt]{1.0\textwidth}{0.3pt}%
       \setlength{\leftmargini}{0em}%
       \setlength{\leftmarginii}{1.5em}%
       \setlength{\leftmarginiii}{1.5em}%
       \setlength{\leftmarginiv}{1.5em}%
       \setlength{\leftmarginv}{1.5em}%
       \begin{description}}
      {\end{description}\medskip
       \end{minipage}
       }

\long\def\comment#1{}

\long\def\symbolfootnote[#1]#2{\begingroup%
\def\thefootnote{\fnsymbol{footnote}}\footnote[#1]{#2}\endgroup}

\ifCLASSINFOpdf
\else
\fi
\begin{document}
%
\title{WiNV: A Framework for Web-based Interactive Scalable Network Visualization}

\author{\IEEEauthorblockN{Hassan Gobjuka}
\IEEEauthorblockA{Verizon\\
919 Hidden Ridge\\
Irving, TX 75083\\
Email: hasan.gobjuka@verizon.com}
}

\author{\begin{tabular}[t]{c@{\extracolsep{1em}}c@{\extracolsep{1em}}c}
           Hassan Gobjuka & Kamal A. Ahmat\\
\it        Verizon & \it Department of Information Technology \\
\it        919 Hidden Ridge & \it City University of New York \\
\it        Irving, TX 75038 & \it New York, NY 11101 \\
\it        {\small\tt hasan.gobjuka@verizon.com} & \it {\small\tt kamal.ahmat@live.lagcc.cuny.edu}
\end{tabular}}


%


\maketitle

\begin{abstract}
In this paper we introduce WiNV -– A framework for web-based interactive scalable network visualization. WiNV enables a new class of rich and scalable interactive cross-platform capabilities for visualizing large-scale networks natively in a user''s browser. Extensive experiments show that our system can visualize networks that consist of tens of thousands of nodes while maintaining fast, high-quality interaction.
\end{abstract}


%
\IEEEpeerreviewmaketitle

\section{Introduction}
This paper presents a new state-of-the-art network visualization framework which supports user interactions with large-scale networks in a web browser without the need for plug-ins or special-purpose runtime systems. Our framework supports the standard visual information browsing functionalities that include overviewing, zooming, and editing. The WiNV framework supports information discovery in two modes; the Standalone Mode and Google Earth Mode. In the Standalone mode, the application runs as a native application. In the Google Earth mode, WiNV visualizes the network on Google Earth enabling the user to get better understanding of wide-spread networks. Interaction is used to mold the network layout into the user''s own mental model and for editing the information being visualized such as device configurations or network layout, if the underlying topology discovery algorithm is unable to discover some connections.

\section{Presentation}
\label{model}

\subsection{Background}

Visualization of large scale graphs have been addressed in the research community \cite{Auber} and \cite{Herman}; open source community \cite{Pajek,jung,Cytoscape}; and industry \cite{TomSawyer}. However, to our knowledge, most of the tools developed so far are general purpose in such that they are not capable to visualize networks at the detailed level may be required by users. Furthermore, none of these tools is web-based, and thus, they don't benefit from the advantages the web offers. We have developed a web-based extensible framework which enables interactive visualization of networks that may consist of hundreds of thousands of devices natively in a web browser. Our framework was developed initially as a presentation-tier for the Network Management System has been developed in \cite{hg2010} and \cite{hbton}.\\
In multi-tiered web-based frameworks, presentation tiers (i.e. clients) are classified into either thick or thin client. A thick client typically provides most of the functionality needed for rendering the data in the browser. Thus, most of the processing time is done in the client side and the interaction with the back-end is relatively low. In thin-clients, on the other hand, most of the processing is done on the back-end and consequently most of the front-end resources can be used for presentation purposes, which in turn achieves higher scalability. However, this approach requires constant communication with the back-end. Web-based presentation-tier applications can also be classified into native or plug-in applications. Examples of plug-ins that are used for browser applications are Adobe Flash and Microsoft Silverlight. The functionality of plug-in-based applications is restricted by the capabilities of the base plug-in. Native applications don't require any application to be pre-installed and, consequently, they can utilize all the features provided by technologies used such as DHTML, JSP, AJAX, and DWR \cite{dwr}. We chose to develop WiNV as thin-client native application but with the capability of integrating it with other native frameworks such as Google Earth.

\subsection{Architecture}

As we stated earlier, the main contribution of this work is a scalable web-based framework for visualizing very large networks while providing smooth interaction and editing capability in a user''s web-browser. The framework architecture is shown in Figure \ref{arch}. To make WiNV flexible and lightweight, the architecture has been designed so that the minimal amount of client's resources are consumed. Furthermore, WiNV can run in any basic browser with no external plug-ins.
WiNV supports two modes: Standalone mode and Google Earth mode. In both modes, most of computation including layout, interaction and editing algorithms run on the back-end while the client only renders the network. Network elements such as devices, hosts and links are generated in-the-fly through the AJAX-based DWR technology and don't require any web-page refreshing if any changes are made, giving a very smooth interaction experience. WiNV framework consists of the following tiers:

\begin{itemize}
  \item \textbf{Presentation Tier} represents a web-browser running on a client machine. Besides visualizing the network, the browser reacts to events triggered by the user (e.g through the mouse) such as clicking on a network segment or creating a network connection between two devices.
  \item \textbf{Network Tier} represents the communication between the presentation and server tiers.
  \item \textbf{Server Tier} represents the ``engine" of the framework. All algorithms used for layout and other computations are loaded within the server tier.
  \item \textbf{Persistent Tier} is used to interact with the database. All changes made to the network layout by the user are reflected to the persistent tier.
\end{itemize}

The WiNV architecture is modular and provides a set of extendable APIs that can be used by any web-based application to visualize the network. The set of APIs can be also used for plugging in multiple different layout algorithms. These algorithms then can be selected through the user interface.

\section{Demonstration}
\filefig{The scalable modular architecture used in WiNV.}
{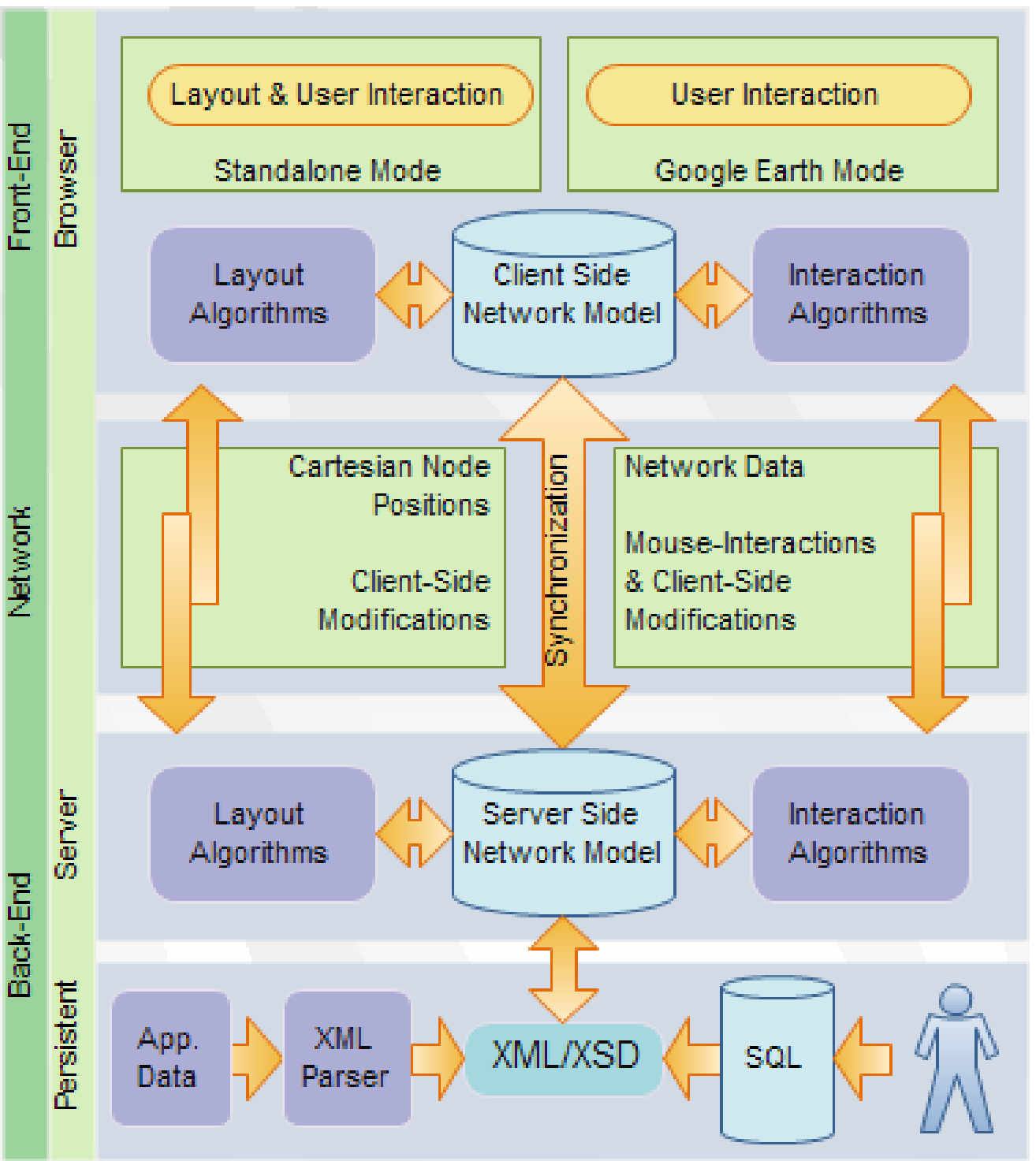}    
{arch}      
{3.1in}        

The framework can receive the data from an XML file that contains network representation, or it can be integrated to exchange the date with the middle tier of a web-based application that implements the framework's APIs. The demo phases consist of (1) Rendering - Displaying the network layout based on the initial data coming from the backend or after a change has been made; and (2) Interaction - Capturing the user command and computing modifications to the network. This process includes, for instance, zooming in, or out, changing device and link configurations or editing the topology. If the framework is set up in Google Earth mode, the user can view the network initially in very high level and a detailed layout will be rendered when zooming in the area that has the target network segment. At the very detailed level, the user could view network device interfaces and direction interconnections among them (i.e. physical network topology.) The user can also view VLAN and ISP level configurations and device/link statistics. Figure \ref{system2} demonstrates high-level network visualization in Google Earth mode. In Standalone mode, the networks are usually grouped based on their IP prefixes or physical location (i.e IP address longitude and latitude) and each group is displayed as a cloud. When the user clicks on a cloud, it expands and that network is rendered at more detailed level. At the most detailed level, all devices, hosts, and peripherals in that segment are rendered. The number of levels, data refreshing rate, and grouping methods are configurable and can be specified by the user.


\filefig {A screenshot of WiNV displaying high-level network layout in Google Earth mode.}
{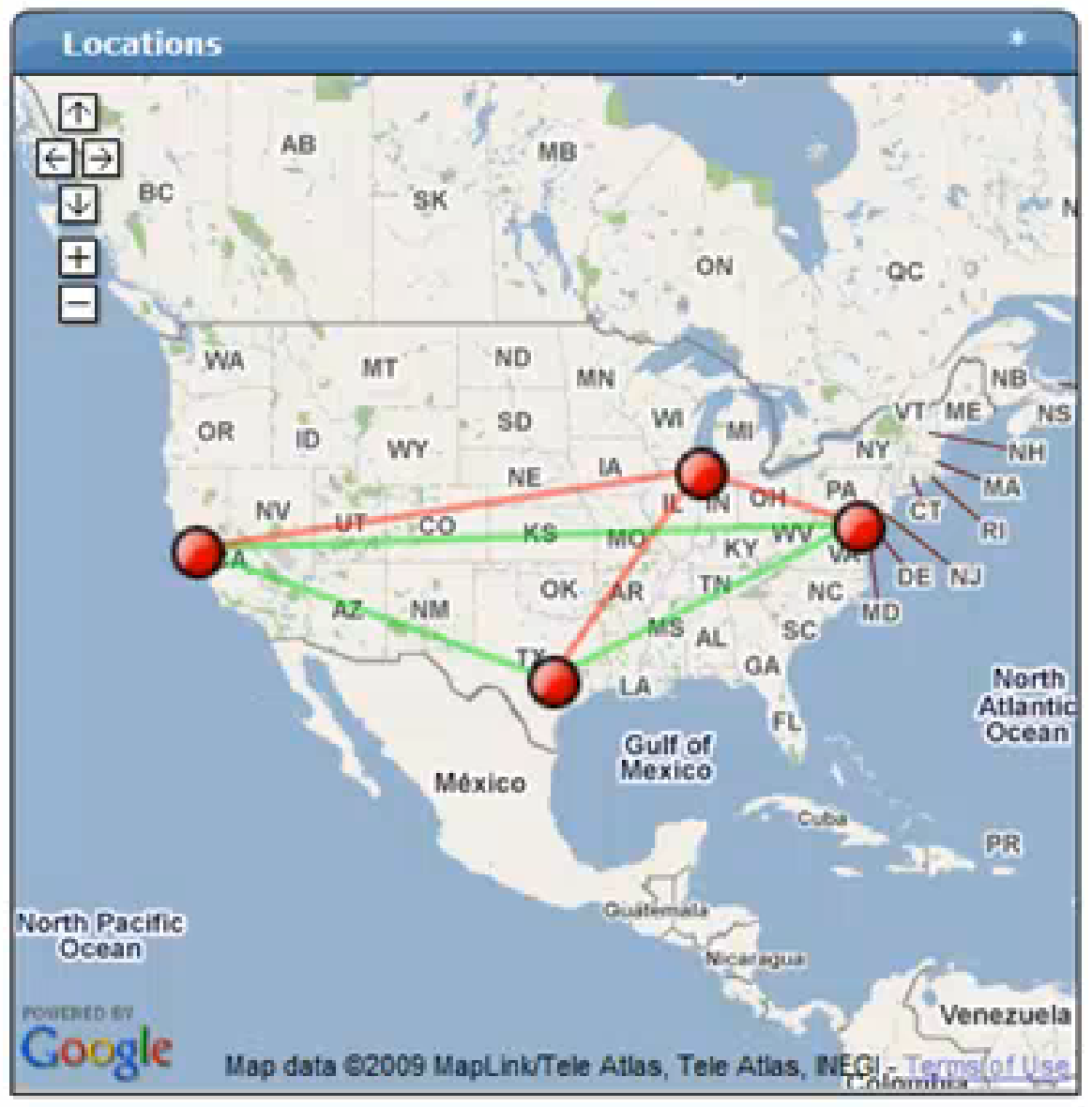}    
{system2}      
{2.7in}        





\begin{thebibliography}{1}

\bibitem{Auber}
D. Auber. Tulip. In P. Mutzel, M. J¨unger, and S. Leipert,
\newblock {\em editors, 9th Symp. Graph. Drawing},
\newblock volume 2265 of Lecture Notes in Computer Science, pages 335–337. Springer-Verlag, 2001.

\bibitem{Pajek}
V. Batagelj, and A. Mrvar,
\newblock {\em Pajek - program for large network analysis},
\newblock Connections, 21:47–57, 1998.


\bibitem{dwr}
Direct Web Remoting,
\newblock {\em Open source online application},
\newblock available at http://directwebremoting.org/dwr/index.html.

\bibitem{hg2010}
H. Gobjuka,
\newblock {\em Topology Discovery for Virtual Local Area Networks},
\newblock in Proc. IEEE INFOCOM Mini-Conference 2010.

\bibitem{hbton}
H. Gobjuka, and Y. Breitbart,
\newblock {\em Ethernet Topology Discovery for Networks with Incomplete Information},
\newblock IEEE/ACM Transactions on Networking, 2010, In Press.

\bibitem{Herman}
I. Herman, G. Melançon, and M. S. Marshall,
\newblock {\em Graph visualization and navigation in information visualization: A survey },
\newblock IEEE Transactions on Visualization and Computer Graphics, 6(1):24-43, 2000.

\bibitem{jung}
J. O'Madadhain, D. Fisher, S. White, and Y. Boey,
\newblock {\em The JUNG (Java Universal Network/Graph) Framework},
\newblock Technical Report UCI-ICS 03-17, School of Information and Computer Science
University of California, Irvine.

\bibitem{Cytoscape}
P. Shannon, A. Markiel, O. Ozier, N. S. Baliga, J. T.Wang, D. Ramage, N. Amin, B. Schwikowski, and T. Ideker,
\newblock {\em Cytoscape: a software environment for integrated models of biomolecular interaction networks},
\newblock Genome Res, 13(11):2498–2504, November 2003.

\bibitem{TomSawyer}
Tom Sawyer Software,
\newblock Tom sawyer visualization, 2009.


\end{thebibliography}
%

\end{document}